# Advanced micropillar cavities: room-temperature operation of microlasers

Andrey Babichev,[1,*] Alexey Blokhin,[1] Yuriy Zadiranov,[1] Yulia Salii,[1] Marina Kulagina,[1] Mikhail Bobrov,[1] Alexey Vasil'ev,[1] Sergey Blokhin,[1] Nikolay Maleev,[1] Ivan Makhov,[2] Natalia Kryzhanovskaya,[2] Leonid Karachinsky,[3] Innokenty Novikov,[3] and Anton Egorov[3]
[1]Ioffe Institute, Saint Petersburg, Russia
[2]HSE University, Saint Petersburg, Russia
[3]ITMO University, Saint Petersburg, Russia

*a.babichev@mail.ioffe.ru

Abstract
High-quality micropillar cavities were grown using molecular-beam epitaxy. Stable continuous-wave lasing at room-temperature was demonstrated for microlasers with semiconductor and hybrid output mirrors. At 300 K, single-mode lasing was demonstrated for micropillars with a diameter of 5 μm at a wavelength of 960 nm, with a minimum lasing threshold of ~1.2 mW and the quality-factor at threshold exceeding 8000.

Quantum dot (QD) based microlasers are implemented in various configurations: microdisk,[1–3] racetrack,[4] ultra-short stripe,[5] and vertical microcavities.[6–8]

Ultrafast spin lasers can be realized in micropillar cavity geometry.[9,10] This concept may also be of interest for neuromorphic computing (NC) and quantum nanophotonics.[11,12] Reservoir computing (RC) mitigates the need for intensive backpropagation through time training, simplifying the use of recurrent NC.[13] An array of diffractively-coupled (DC) lasers is one implementation of optical RC.[14] Optically pumped micropillar lasers[12] or vertical-cavity surface-emitting laser (VCSELs)[15] can be used as the type of DC lasers. The advantage of DC VCSELs is the ability to lasing at room temperature, but the large pitch of commercially available VCSELs (250 μm) limits the reservoir (a network with fixed recurrent connections) size.[16] The problem of large pitch can be partially addressed with custom-designed VCSELs,[17] but the reservoir consists of only of 24 nodes[15] and requires further scaling.

The ability to produce an ultra-dense array of DC lasers is the main benefit of using optically pumped micropillar lasers,[12,18] whose pitch is approximately ten times smaller (8 μm) compared to their VCSEL counterpart.[15] The major drawback of optically pumped micropillar cavity lasers is their low operating temperature.[16,19,20] In addition, low power-conversion efficiency (PCE) is also discussed.[20,21] The latter value for micropillar lasers can be improved by using low-absorbing (at the pump laser wavelength) semiconductor,[6,19,22] dielectric,[23] and hybrid[7] mirrors. Despite the increase in the PCE value, the maximum operating temperature of micropillar lasers does not exceed 220 K.[7,22]

In this Letter, we report a demonstration of lasing at room temperature in micropillar cavity lasers using optical pumping.

Firstly, numerical simulations were carried out for a 5 μm diameter micropillar cavity in designs with different output mirrors to realize high quality microcavity. Details of the modelling can be found in Ref. 7. To qualitatively compare the various designs, a simplified model of a cold vertical microcavity was used. The first one-lambda cavity is made entirely of semiconductors and includes 37.5 and 32.0 pairs of $Al_{0.2}Ga_{0.8}As/Al_{0.9}Ga_{0.1}As$ layers in the bottom and top distributed Bragg reflectors (DBRs). A full description of this structure is given below. Increasing the temperature from 77 to 300 K leads to an enlargement in the quality-factor (*Q*-factor) from 66500 to 82400. The temperature behavior of the *Q*-factor is determined by the temperature dependence of the refractive index and extinction coefficient for AlGaAs layers. Using a design with a hybrid output mirror (with two pairs of $SiO_2/Ta_2O_5$) results in an increase in the *Q*-factor to 122000and 152700 at 77 and 300 K. As a result, in the examined design with a semiconductor and hybrid output mirror, the *Q*-factor can be increased at least twofold compared to the previously created microcavities based on low-absorbing mirrors and grown molecular-beam epitaxy (MBE).[6,7]

The structure of the one-lambda GaAs cavity was grown by MBE. Growth was performed on a semi-insulating GaAs wafer. The bottom DBR included 37.5 pairs of alternating $Al_{0.2}Ga_{0.8}As/Al_{0.9}Ga_{0.1}As$. The gain region, based on InGaAs QDs, is positioned in the center of the cavity. Self-organized QDs were formed at lower temperatures using the Stranski-Krastanov growth mode. The gain region consisted of three layers of QDs stacked on top of each other. Details of the gain region optimization are presented in Ref. 24. The initial QDs were used to blueshift the position of the ground-state peak in the photoluminescence spectrum, which is ~1030 nm at 300 K. The growth conditions were modified to increase the gain of the QDs layers by exception the vertical alignment of QDs (the position of the next layer of QDs above the previous one). The top DBR included 32 pairs of $Al_{0.2}Ga_{0.8}As/Al_{0.9}Ga_{0.1}As$ layers.

Micropillar lasers were fabricated using a dry etching process. A reflowed-photoresist was used as a hard mask for dry etching to a depth of approximately 10 μm. To create the hybrid output mirror, two pairs of $SiO_2$/ $Ta_2O_5$ were deposited using magnetron sputtering.[7]

A Cryostation® s50 closed-cycle optical cryostat (Montana Instruments) was used to study the samples. A semiconductor laser diode with a wavelength of 808 nm (in continuous-wave mode) was used for optical pumping of the micropillars. Typically, the focused spot of optically pumped micropillar lasers is close to the limit of a 100× objective (about 1–2 µm),[6,12,19] which is associated with the study of high-beta-emitting microcavities (with typical diameters of several micrometers[25–27]) and single-photon sources.[28,29] Recently, A. Babichev *et al.* used full surface pumping of the micropillar, which made it possible to obtain whispering-gallery modes lasing in the thermoelectrical cooling range (up to 170 K).[6] Here, we also applied full surface pumping of the micropillars to minimize the heating effect, which limits the maximum lasing temperature of the micropillar laser with a pump spot of about 1–2 µm. A 20× objective (Mitutoyo M Plan Apo NIR) with a numerical aperture of 0.42 was used for pumping and collecting luminescence. An SR-500i spectrometer (Andor Shamrock) with a focal length of 500 mm and a resolution of 0.05 nm (with a grating of 1200 lines/mm) was used to collect emission. A thermoelectrically cooled, back-illuminated silicon CCD matrix DU 401A BVF (Andor Shamrock) was combined with a monochromator.

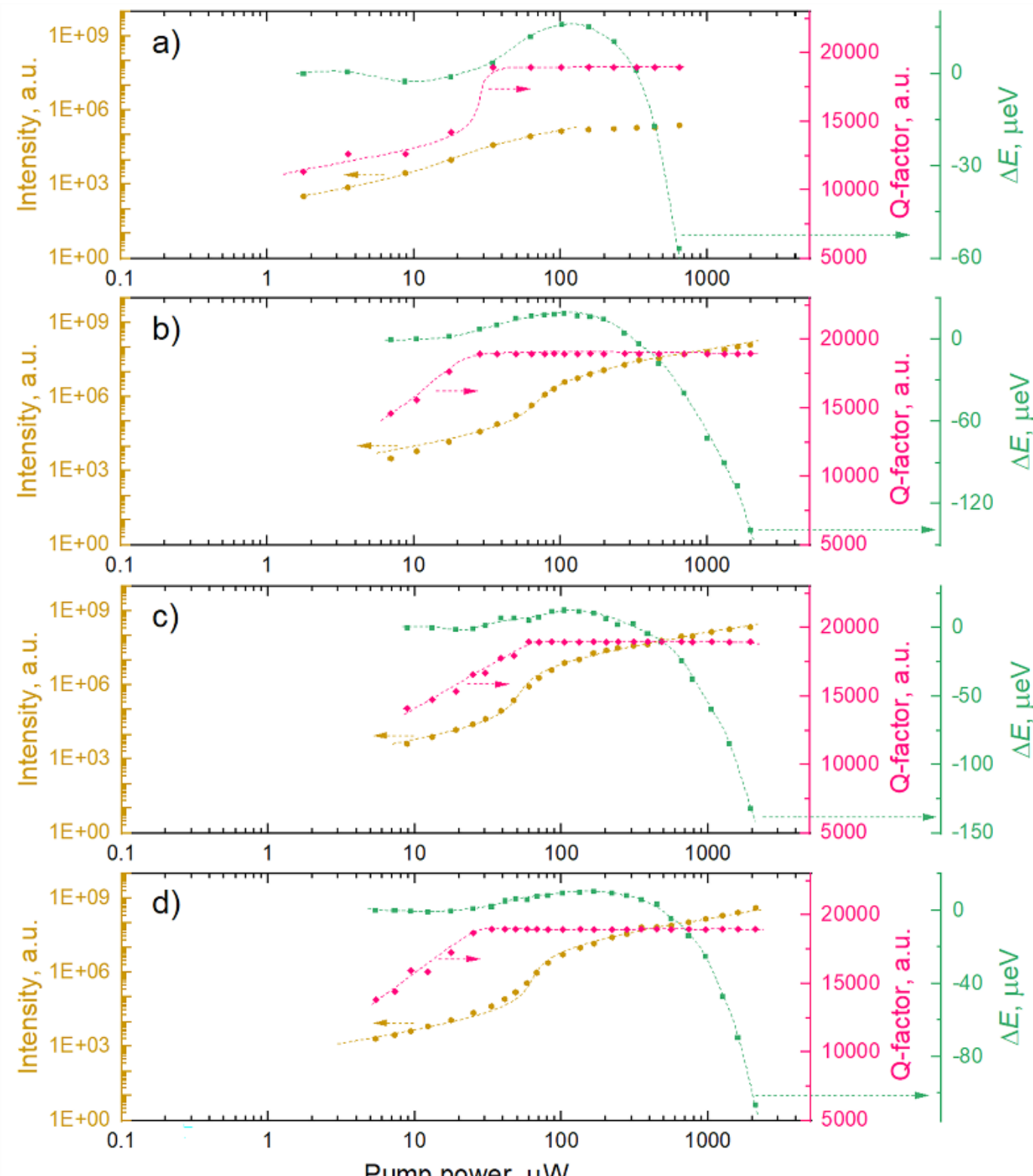

Fig. 1. I-O characteristics (left Y-axes) of micropillar lasers at 77 K. Panels (a), (b), (c) and (d) correspond to pillar diameters of 3.5, 4.0, 4.5 and 5.0 µm. The right Y-axes show the dependence of the Q-factor and the mode energy shift on the pump power extracted at 77 K.

We first present the results of input-output (*I-O*) measurements of micropillar lasers at 77 K. The optical characteristics of the micropillar laser with a hybrid output mirror for diameters of 3.5, 4.0, 4.5 and 5.0 µm are shown in Fig. 1. To determine the integrated intensity of the luminescence peak and its linewidth, a full Voigt line shape approximation was used, taking into account laser-related effects reflected in the spectral lineshape.[23] This profile is better suited for the analysis of luminescence spectra than the pseudo-Voigt profile,[19] since it allows the Lorentzian and Gaussian components of the linewidth to be obtained separately.[8,30,31] The *I-O* curves have an *S*-type dependence on a double logarithmic scale. With increasing pump power, a narrowing of the linewidth (limited by the spectral resolution of the monochromator) is observed for all examined pillar diameters (see Fig. 1). To determine the lasing threshold and *Q*-factor at threshold, the rate-equations fitting[19] was used. The pump power ($P_{pump}$) can be determined by the expression[19]:

$$P_{pump}=A\gamma[BP_{out}(1+\xi+\beta(BP_{out}-\xi)/(1+BP_{out}))]/\beta,$$

where $A$ and $B$ are the scaling factors, $\gamma$ is the cavity decay rate, $P_{out}$ is the output power, and $\beta$ is the coupling factor between the lasing mode and the QDs gain. The factor $\xi$ is the mean number of spontaneous emission photons in the cavity at transparency and can be estimated using the expression: $\xi=n_0\beta/\gamma\tau_{sp}$, where $n_0$ is the number of excitons at transparency and $\tau_{sp}$ is the spontaneous exiton lifetime. The value of $n_0$ can be estimated by multiplying the pillar diameter by the QDs density. The value of $\gamma$ is derived from the expression: $\gamma=2\pi\nu_L/Q$, where $\nu_L$ is the lasing wavelength and $Q$ is the quality-factor at threshold. The lasing threshold $P_{th}$ can be determined from the dependence $P_{pump}(P_{out})$ using the reduced expression:

$$P_{th}=A\gamma[\xi(1-\beta)+1+\beta)]/2\beta.$$

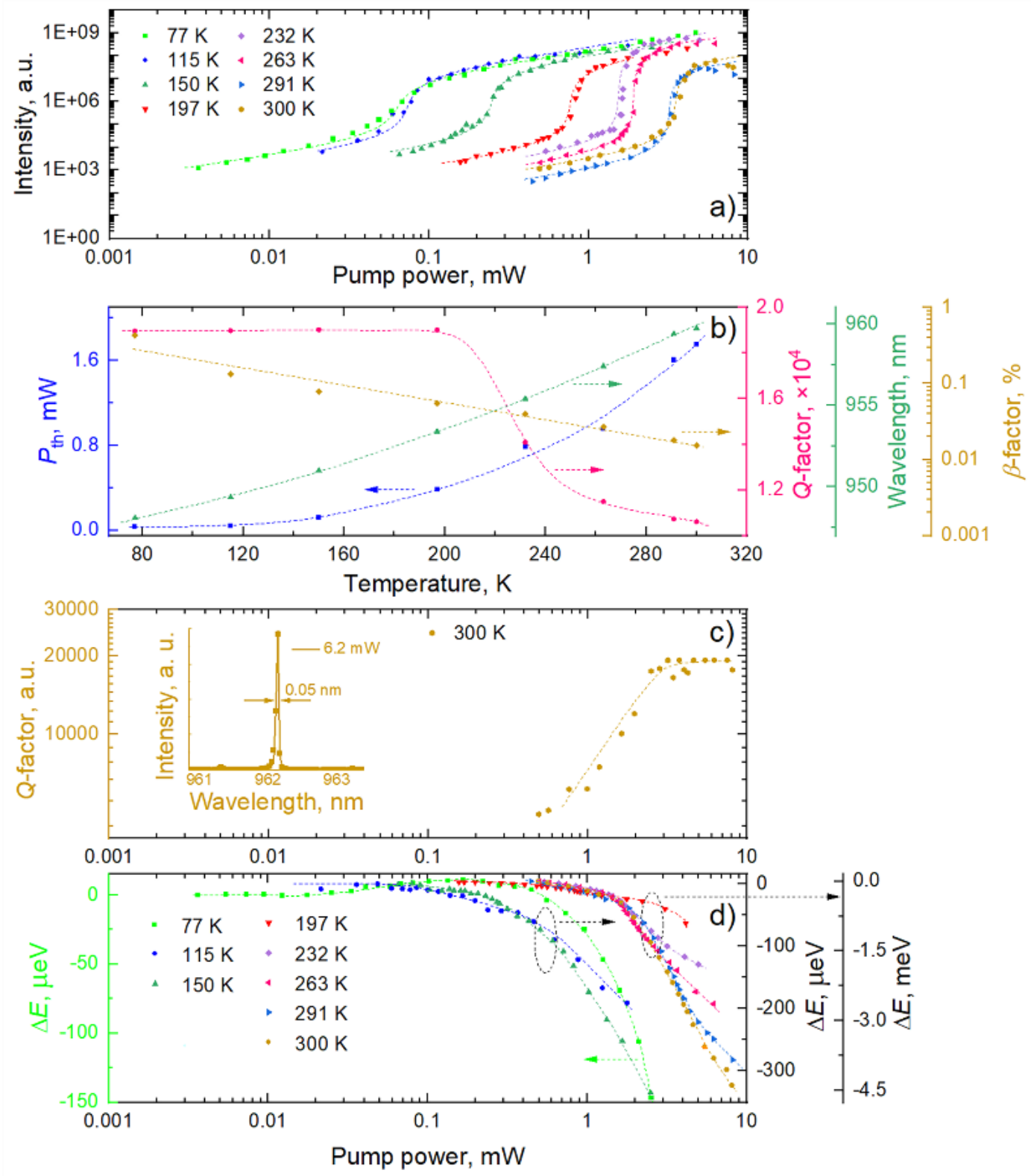

Fig. 2. (a) I-O characteristics of the micropillar laser with a diameter of 5.0 µm determined in the temperature range 77–300 K; (b) Lasing threshold (left Y-axis), Q-factor at threshold, β-factor and threshold wavelength (right Y-axes) extracted at different temperatures; (c) Pump-power-dependent Q-factor at 300 K. The inset shows the lasing spectra at 6.2 mW pump power; (d) Mode energy shift as a function of pump power at 77 K (left Y-axis), 115, 150 K (right Y-axis), and at 197, 232, 263, 291 and 300 K (right Y-axis).

For a pillar with a diameter of 3.5 µm, a slightly superlinear dependence of pump power is shown. The linewidth narrowing with pump power is about 47 µeV (from 116 to 69 µeV) and is limited by the spectral resolution of the monochromator (see Fig. 1(a)). The extracted *β*-factor value is (11.5±3.6) %, which leads to a lasing threshold of (9.5±0.2) µW. Further evidence of ongoing lasing transition would be required using second-order autocorrelation measurements,[22] which are commonly used in the analysis of ultra-high-β microcavities.[26,30,32]

For a pillar with a diameter of 4.0 µm, the lasing threshold is (26.2±0.6) µW (see Fig. 1(b)). A similar threshold power is extracted for 4.5 µm pillar ((25.4±0.6) µW). For a pillar with a diameter of 5.0 µm, the lasing threshold increases to (31.3±0.7) µW. These values are close to the extremely low lasing thresholds (30–40 µW for 2.9–5.4 µm pillars[19]).

For the examined pillar diameters, the *Q*-factor at threshold is limited by the spectral resolution of the monochromator and is at least 19000. Apart from an increase in the *Q*-factor at threshold for a 5.0 µm pillar (from 15000[19] to at least 19000), no inhomogeneous linewidth broadening was demonstrated with increasing pump power (see Fig. 1(d)).

To study the thermal effects, the mode energy shift (Δ*E*) is analyzed as a function of the pump power (see Fig. 1). At low pump power, a slight mode redshift (–2.7 µeV) is demonstrated for a 3.5 µm pillar, which is associated to the bandgap shrinkage.[19] A further increase in pump power leads to a blue shift of the mode (16 µeV at 10.7 $P_{th}$), which can be associated with the band filling and the plasma effect.[19] The predominance of thermal effects over plasma ones is noticeably larger than 11 $P_{th}$ (see Fig. 1(a)). It has been previously shown that increasing the pillar diameter from 2.9 to 5.4 µm leads to a decrease in thermal shift.[19] Similar behavior is observed for the examined microlasers (see Fig. 1(d)). To minimize thermal effects, a pillar laser with a diameter of 5.0 µm was chosen for temperature studies. In the temperature range of 77–300 K, the *I-O* curves of the 5 µm pillar laser with a hybrid output mirror exhibit an *S*-shaped dependence on a double logarithmic scale (see Fig. 2(a)). This fact, as well as the narrowing of the emission linewidth, confirm the transition to lasing.[19,22] The fitting results are shown in Fig. 2(a). The lasing threshold and the *Q*- factor at threshold extracted at different temperatures are shown in Fig. 2(b). Increasing the temperature from 77 to 300 K leads to an enlarge in the lasing threshold from (31.3±0.7) µW to (1.75±0.14) mW. The spontaneous emission factor is (0.4±0.1) % at 77 K and decreases to (0.018±0.0006) % with an increase in temperature to 300 K (see Fig. 2(b)).

Although at low temperatures the *Q*-factor at threshold is limited by the monochromator resolution (at least 19000), increasing the temperature to 300 K leads to a *Q*-factor of (10600±710). The decrease in the *Q*-factor with increasing temperature in the experiment may be due to absorption in the gain region. The temperature dependence of the refractive index and extinction coefficient of the gain region has a significant effect on the *Q*-factor depending on the temperature. The absorption of GaAs (material of the cavity) at the pump wavelength (808 nm) increases significantly with temperature. Furthermore, as the temperature rises, the ground state of QDs rapidly shifts toward longer wavelengths compared to the resonance energy (wavelength) of the microcavity. This means that excited states and, ultimately, the wetting layer begin to influence the imaginary part of the complex refractive index (absorption coefficient). As a result, the greater the absorption, the lower the *Q*-factor, as observed experimentally.

The pump-power-dependent *Q*-factor extracted at 300 K is shown in Fig. 2(c). There is no decrease in the *Q*-factor at high powers. This fact, along with the clearly defined *S*-shaped I-O characteristic and the narrowing of the linewidth with increasing the pump power (see Fig. 2(c)), convincingly indicates a transition to the coherent emission regime.[19,22] A typical lasing spectrum is displayed in the inset of the Fig. 2(c). The position of the wavelength at the threshold power extracted at different temperatures is shown in Fig. 2(b). At a temperature of 77 K, the lasing wavelength, determined just above the threshold, is 948.0 nm. Increasing the temperature to 300 K leads to a shift in the emission wavelength to 959.6 nm.

The mode energy shift extracted at different temperatures is shown in Fig. 2(d). At a pump power of 2 mW, increasing the temperature from 77 to 150 K leads to an enlarge in the mode energy shift from –98 to –290 μeV. At a temperature of 300 K, the mode energy shift is –775 μeV (at 2 mW pump power).

Figure 3(a) shows the results of a 5 μm pillar laser with a hybrid output mirror, the position of which is shifted by 920 μm (the position of the pillar on the substrate) compared to the pillar, the results of which are presented in Fig. 2. At 300 K, the *I-O* curve has an *S*-shape (see Fig. 3(a)), which, along with the narrowing of the emission linewidth (without its broadening at high pump power), confirms the laser transition._The lasing threshold and the *Q*-factor at threshold are (1.73±0.14) mW and (10400±530) (see Fig. 3(a)), which are close to those discussed above. The mode energy, determined just above the threshold, is shifted by 98 μeV compared to the pillar, the results of which are shown in Fig. 2. Increasing the pump power to 2.8 mW leads to a *Q*-factor of more than 19000. This value is maintained up to a pump power of 8 mW.

In addition to studying microlasers with a hybrid output mirror, 5 μm pillar lasers with a semiconductor output mirror were investigated. At 300 K, the *I-O* curve has an *S*-shape (see Fig. 3(b)). The fitting results are shown in Fig. 3(b). The lasing threshold and *Q*-factor at threshold are (1.43±0.12) mW and (8900±440), which are lower than those of pillars with a hybrid output mirror. Although a narrowing of the emission linewidth is observed, a broadening is also established at high pump power. Additional quantum optical measurements can be used as further evidence for the transition to coherent lasing.[22]

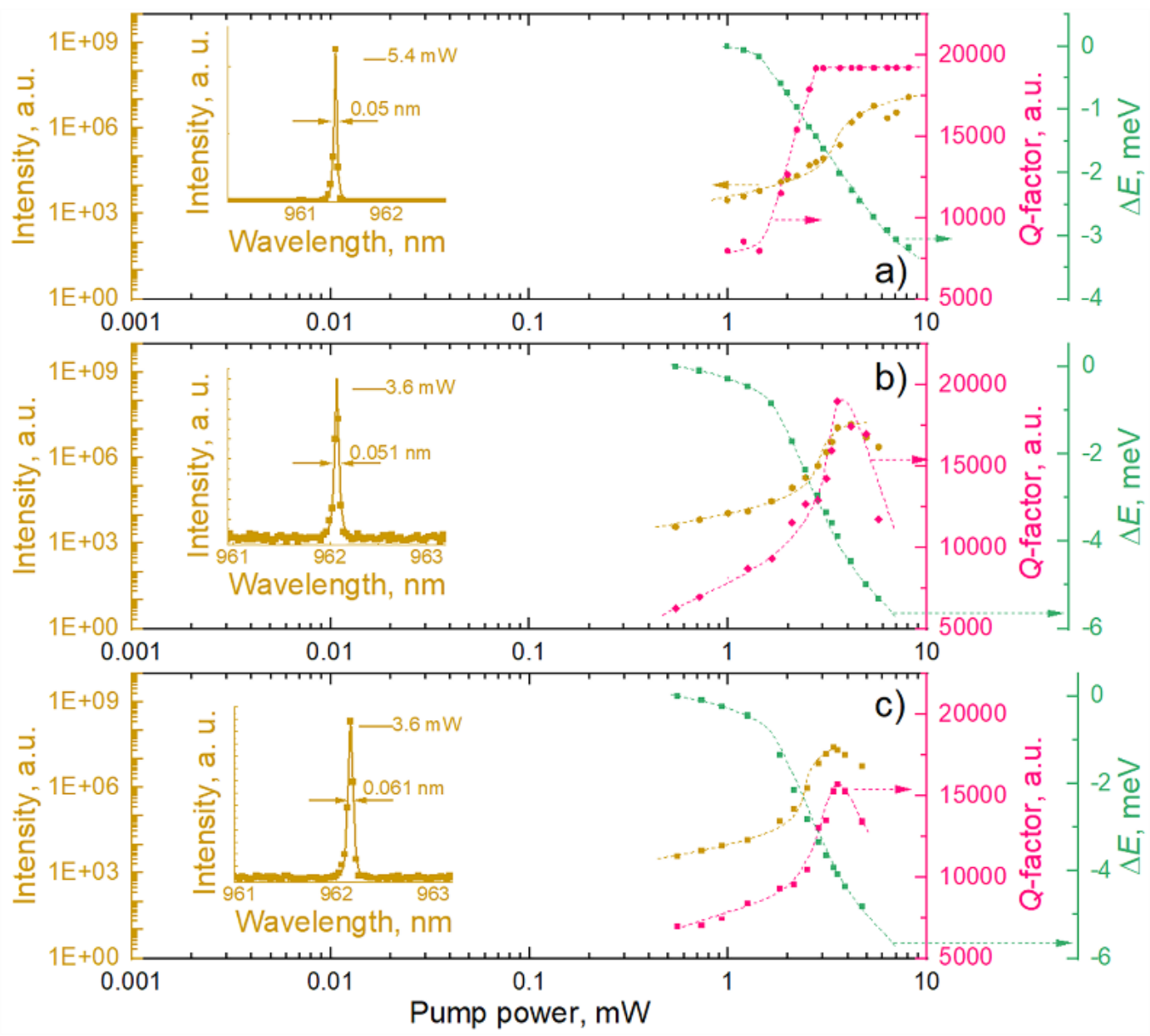

Fig. 3. (a) I-O characteristic (left Y-axes) of micropillar laser with a diameter of 5.0 μm determined at 300 K. The right Y-axes show the dependence of the Q-factor and the mode energy shift on the pump power extracted at 300 K. Panels (a) and (b, c) correspond to a micropillar laser with hybrid and semiconductor output mirrors, respectively. The inset to panels (a) and (b, c) shows the lasing spectra at a pump power of 5.4 and 3.6 mW.

Typically, high losses and lower *Q*-factor lead to a decrease in the Purcell factor of micropillar cavities with semiconductor mirrors.[27] A similar situation is observed in VCSELs, where increasing losses in mirrors leads to an increase in the lasing threshold.[33–35] On the contrary, in the experiment a decrease in the threshold is observed along with a reduction in the *Q*-factor from 10400 to 8900. To describe this behavior, it should be noted that the pillars use two types of output mirrors, which differ not only in the number of semiconductor pairs, but also in materials. The use of a hybrid mirror changes the transmission at the pump wavelength (essentially, the PCE[20]), due to two factors: it can increase the mirror reflectivity at the pump wavelength and increases the output mirror losses.[36,37] Although the absorption of $Ta_2O_5$ and $SiO_2$ is 0.02 $\mu m^{-1}$ at 808 nm,[38,39] which is approximately twice that of $Al_{0.2}Ga_{0.8}As$,[40] its impact on the PCE is minimal due to the small overall thickness of the $Ta_2O_5$ layers ($\Delta$PCE ~ 0.12%). The insignificant influence of absorption in the $Ta_2O_5$ and $SiO_2$ layers (at the resonance wavelength) is also confirmed by the larger *Q*-factor compared to the pillar with a semiconductor output mirror.

As shown previously,[20,22] compared to conventional DBR (based on $GaAs/Al_{0.9}Ga_{0.1}As$), the use of a modified design (based on low-absorbing $Al_{0.2}Ga_{0.8}As/Al_{0.9}Ga_{0.1}As$) results in clear fluctuations in reflectivity below the stopband position (clear fringes). Switching from a microcavity with a hybrid mirror to a microcavity with a semiconductor top mirror changes the output mirror reflectivity to ~ 0.12 (due to its oscillation) at the pump wavelength, which leads to a decrease in PCE to approximately 3%. This may be the main reason for the lower lasing threshold for micropillar with a semiconductor output mirror compared to micropillars with a hybrid output mirror.

Figure 3(b) reveals that at high pump power, a rapid decrease in emission intensity is observed for a pillar with a semiconductor output mirror, which may be due to thermal effects. The effect of decreasing the *Q*-factor at threshold on the thermal properties is analyzed. At a pump power of 2 mW, an increase in the mode energy shift to –1.7 meV was demonstrated. The latter value is more than twice as high as the similar value for the pillar with a hybrid output mirror (–0.78 meV, see Fig. 2(d)).

Figure 3(c) shows the result of the study of a 5 µm pillar laser with a semiconductor output mirror whose position is shifted by 920 µm compared to the pillar, the results of which are displayed in Fig. 3(b). The lasing threshold and the Q-factor at threshold are (1.22±0.10) mW and (8100±400). The mode energy, determined just above the threshold, is shifted to 106 µeV compared to the pillar, the results of which are shown in Fig. 3(b). This fact correlates with a small spectral nonuniformity along the radius of 2-inch wafers, which is almost 7 times smaller[41] than the results for the structure grown by the MOVPE method.[12]

In conclusion, microlasers based on micropillar cavity designs were fabricated and analyzed. Low-threshold lasing was realized at 77 K for micropillar lasers with a hybrid output mirror and diameters of 3.5, 4.0, 4.5 and 5.0 µm. The lasing thresholds of the examined micropillars are comparable to the extremely low lasing thresholds discussed earlier.[19]

For the first time, operation of micropillar cavity lasers at room temperature has been demonstrated. This is primarily due to the optimized design of mirrors with large PCE, first invented by S. Reitzenstein *et al.*,[19] and an efficient gain region based on the initial QDs grown by MBE. Further optimization will involve pulsed pumping of microlasers[26] to minimize thermal issues.

Recently, room temperature operation for quasi-planar geometry of photonic-defect cavity lasers[22,23,42] with a lasing threshold of about 4.9 mW for a buried mesa diameter of 5 µm is reported.[22]

For $Ta_2O_5/SiO_2$ and $Si_3N_4/SiO_2$ mirrors, a moderate refractive index contrast ($\Delta n$) is observed ($\Delta n$ = 0.64 and 0.56, respectively). Further research should be aimed at finding dielectrics that are most suitable for the fabrication of mirrors used in micropillar and photonic-defect cavity lasers. Conventional dielecric mirrors suitable for long-wavelength VCSELs include $SiN_y/SiO_x$, $SiC_z/SiO_x$, a-Si/$Al_2O_3$, a-Si/$CaF_2$, a-Si/MgF, $ZnS/CaF_2$ and $ZnS/AlF_3$.[37,43] The extremely large absorption of $SiC_z$ does not allow its use at a wavelength of 808 nm. Compared to the $Ta_2O_5$[38] which acts as a high-index material, the absorption of a-Si is approximately 5 times higher at 808 nm, despite having more than twice the refractive index.[44] In contrast to $Ta_2O_5$, ZnS exhibits approximately 10 times lower absorption.[45] In addition, $Si_3N_4$ does not absorb at 808 nm.[22] Compared to $SiO_2$, which acts as a low-index material, approximately 3 times lower absorption was observed for $MgF_2$[46] and $Al_2O_3$.[47] In addition, $CaF_2$ does not absorb at 808 nm.[48] Another issue, along with ensuring large $\Delta n$ and the absence of absorption, is its thermal resistance.[36] This value for ZnS is 2.8 times smaller compared to a-Si.[36] This fact, along with the large $\Delta n$ value (0.88), makes $ZnS/CaF_2$ an optimal dielectric pair if HCl etching is not required.[36,43] $ZnS/AlF_3$ can be used as an alternative to $ZnS/CaF_2$ due to its larger $\Delta n$ value[49] and process compatibility.[36,43]

The work of A. Babichev, A. Blokhin, A. Vasil'ev, and S. Blokhin was supported by the Russian Science Foundation under Grant 22-19-00221-П, https://rscf.ru/project/22-19-00221/ for the structure design, MBE epitaxy, and the study of PL and lasing spectra. The work of I. Makhov and N. Kryzhanovskaya was supported by the Basic Research Program at the HSE University for support the study of test structures.

## References

[1]A. Zhukov, A. Nadtochiy, A. Karaborchev, N. Fominykh, I. Makhov, K. Ivanov, Y. Guseva, M. Kulagina, S. Blokhin, and N. Kryzhanovskaya, Opt. Lett. **49**, 330 (2024).
[2]E. Moiseev, N. Kryzhanovskaya, M. Maximov, F. Zubov, A. Nadtochiy, M. Kulagina, Y. Zadiranov, N. Kalyuzhnyy, S. Mintairov, and A. Zhukov, Opt. Lett. **43**, 4554 (2018).
[3]F. Zubov, M. Maximov, E. Moiseev, A. Vorobyev, A. Mozharov, Y. Berdnikov, N. Kaluzhnyy, S. Mintairov, M. Kulagina, N. Kryzhanovskaya, and A. Zhukov, Opt. Lett. **46**, 3853 (2021).
[4]I. Makhov, K. Ivanov, E. Moiseev, A. Dragunova, N. Fominykh, N. Kryzhanovskaya, and A. Zhukov, Opt. Lett. **48**, 3515 (2023).
[5]I. Makhov, S. Komarov, N. Fominykh, A. Obraztsova, V. Voitovich, N. Derkach, I. Melnichenko, N. Shandyba, N. Chernenko, M. Solodovnik, A. Lipovskii, Y. Shernyakov, N. Kalyuzhnyy, S. Mintairov, and N. Kryzhanovskaya, Opt. Lett. **50**, 387 (2025).
[6]A. Babichev, I. Makhov, N. Kryzhanovskay, S. Troshkov, Y. Zadiranov, Y. Salii, M. Kulagina, M. Bobrov, A. Vasil'ev, S. Blokhin, N. Maleev, L. Karachinsky, I. Novikov, and A. Egorov, IEEE J. Sel. Top. Quantum Electron. **31**, 1502808 (2025).
[7]A. Babichev, I. Makhov, N. Kryzhanovskaya, A. Blokhin, Y. Zadiranov, Y. Salii, M. Kulagina, M. Bobrov, A. Vasil'ev, S. Blokhin, N. Maleev, M. Tchernycheva, L. Karachinsky, I. Novikov, and A. Egorov, IEEE J. Sel. Top. Quantum Electron. **31**, 1900208 (2025).
[8]I. Limame, C. W. Shih, A. Koulas-Simos, J. Pietsch, L. J. Roche, M. Plattner, A. Koltchanov, S. Rodt, and S. Reitzenstein, Opt. Express **32**, 31819 (2024).
[9]M. Lindemann, G. Xu, T. Pusch, R. Michalzik, M. R. Hofmann, I. Žutić, and N. C. Gerhardt, Nature **568**, 212 (2019).
[10]N. Heermeier, N. Heuser, J. Große, N. Jung, A. Kaganskiy, M. Lindemann, N. C. Gerhardt, M. R. Hofmann, and S. Reitzenstein, Laser Photonics Rev. **16**, 2100585 (2022).
[11]T. Heindel, J. H. Kim, N. Gregersen, A. Rastelli, and S. Reitzenstein, Adv. Opt. Photonics **15**, 613 (2023).
[12]T. Heuser, J. Große, S. Holzinger, M. M. Sommer, and S. Reitzenstein, IEEE J. Sel. Top. Quantum Electron **26**, 1900109 (2019).
[13]A. Skalli, M. Goldmann, N. Haghighi, S. Reitzenstein, J. A. Lott, and D. Brunner, Commun. Phys. **8**, 68 (2025).
[14]D. Brunner, and I. Fischer, Opt. Lett. **40**, 3854 (2015).
[15]M. Pflüger, D. Brunner, T. Heuser, J. A. Lott, S. Reitzenstein, and I. Fischer, Opt. Lett. **49**, 2285 (2024).
[16]T. Heuser, M. Pflüger, I. Fischer, J. A. Lott, D. Brunner, and S. Reitzenstein, JPhys Photonics **2**, 044002 (2020).
[17]M. Pflüger, D. Brunner, T. Heuser, J. A. Lott, S. Reitzenstein, and I. Fischer, Opt. Express **31**, 8704 (2023).
[18]T. Heuser, J. Große, A. Kaganskiy, D. Brunner, and S. Reitzenstein, APL Photonics **3**, 116103 (2018).
[19]C. W. Shih, I. Limame, S. Krüger, C. C. Palekar, A. Koulas-Simos, D. Brunner, and S. Reitzenstein, Appl. Phys. Lett. **122**, 151111 (2023).
[20]L. Andreoli, X. Porte, T. Heuser, J. Große, B. Moeglen-Paget, L. Furfaro, S. Reitzenstein, and D. Brunner, Opt. Express **29**, 9084 (2021).
[21]S. Reitzenstein, A. Bazhenov, A. Gorbunov, C. Hofmann, S. Münch, A. Löffler, M. Kamp, J. P. Reithmaier, V. D. Kulakovskii, and A. Forchel, Appl. Phys. Lett. **89**, 051107 (2006).
[22]K. Gaur, S. Tripathi, F. Laudani, A. Barua, I. Limame, A. Koulas-Simos, S. Rodt, and S. Reitzenstein, Laser Photonics Rev. **19**, e00533 (2025).
[23]K. Gaur, C. W. Shih, I. Limame, A. Koulas-Simos, N. Heermeier, C. C. Palekar, S. Tripathi, S. Rodt, and S. Reitzenstein, Appl. Phys. Lett. **124**, 041104 (2024).
[24]A. Babichev, A. Blokhin, Y. Zadiranov, Y. Salii, M. Kulagina, M. Bobrov, Y. Kovach, A. Vasil'ev, S. Blokhin, N. Maleev, I. Makhov, N. Kryzhanovskaya, M. Tchernycheva, L. Karachinsky, I. Novikov, and A. Egorov, OpticaOpen, 124547 (2025). Available online (dated on 12.29.2025): https://doi.org/10.1364/opticaopen.29598755
[25]M. Lermer, N. Gregersen, M. Lorke, E. Schild, P. Gold, J. Mørk, C. Schneider, A. Forchel, S. Reitzenstein, S. Höfling, and M. Kamp, Appl. Phys. Lett. **102**, 052114 (2013).
[26]S. M. Ulrich, C. Gies, S. Ates, J. Wiersig, S. Reitzenstein, C. Hofmann, A. Löffler, A. Forchel, F. Jahnke, and P. Michler, Phys. Rev. Lett. **98**, 043906 (2007).
[27]H. Deng, G. L. Lippi, J. Mørk, J. Wiersig, S. Reitzenstein, Adv. Opt. Mater. **9**, 2100415 (2021).
[28]I. Limame, P. Ludewig, C.-W. Shih, M. Hohn, C. C. Palekar, W. Stolz, and S. Reitzenstein, Opt. Quantum **2**, 117 (2024).
[29]S. A. Blokhin, M. A. Bobrov, N. A. Maleev, J. N. Donges, L. Bremer, A. A. Blokhin, A. P. Vasil'ev, A. G. Kuzmenkov, E. S. Kolodeznyi, V. A. Shchukin, N. N. Ledentsov, S. Reitzenstein, and V. M. Ustinov, Opt. Express **29**, 6582 (2021).
[30]A. Koulas-Simos, J. Buchgeister, M. L. Drechsler, T. Zhang, K. Laiho, G. Sinatkas, J. Xu, F. Lohof, Q. Kan, R. K. Zhang, F. Jahnke, C. Gies, W. W. Chow, C.-Z. Ning, and S. Reitzenstein, Laser Photonics Rev. **16**, 2200086 (2022).
[31]A. Koulas-Simos, C. C. Palekar, K. Gaur, I. Limame, C.-W. Shih, B. L.T. Rosa, C.-Z. Ning, and S. Reitzenstein, Laser Photonics Rev. **18**, 2400271 (2024), doi: 10.1002/lpor.202400271.
[32]S. Kreinberg, W. W. Chow, J. Wolters, C. Schneider, C. Gies, F. Jahnke, S. Höfling, M. Kamp, and S. Reitzenstein, Light:Sci. Appl. **6**, e17030 (2017).
[33]S. A. Blokhin, A. V. Babichev, A. G. Gladyshev, I. I. Novikov, A. A. Blokhin, M. A. Bobrov, N. A. Maleev, V. V. Andryushkin, D. V. Denisov, K. O. Voropaev, V. M. Ustinov, V. E. Bougrov, A. Y. Egorov, and L. Y. Karachinsky, Opt. Eng. **61**, 096109 (2022).
[34]S. A. Blokhin, M. A. Bobrov, A. A. Blokhin, A. G. Kuzmenkov, N. A. Maleev, V. M. Ustinov, E. S. Kolodeznyi, S. S. Rochas, A. V. Babichev, I. I. Novikov, A. G. Gladyshev, L. Ya. Karachinsky, D. V. Denisov, K. O. Voropaev, A. S. Ionov, and A. Yu. Egorov, Semiconductors **53**, 1104 (2019).
[35]S. A. Blokhin, M. A. Bobrov, A. A. Blokhin, A. G. Kuzmenkov, N. A. Maleev, V. M. Ustinov, E. S. Kolodeznyi, S. S. Rochas, A. V. Babichev, I. I. Novikov, A. G. Gladyshev, L. Ya. Karachinsky, D. V. Denisov, K. O. Voropaev, A. S. Ionov, and A. Yu. Egorov, Opt. Spectrosc. **127**, 140 (2019).
[36]A. V. Babichev, Y. N Kovach, S. A. Blokhin, L. Ya Karachinsky, I. I. Novikov, A. Yu. Egorov, S.-C. Tian, and D. Bimberg, JPhys Photonics **7**, 032001 (2025).
[37]A. Babichev, S. Blokhin, E. Kolodeznyi, L. Karachinsky, I. Novikov, A. Egorov, S.-C. Tian, and D. Bimberg, Photonics **10**, 268 (2023).
[38]T. J. Bright, J. I. Watjen, Z. M. Zhang, C. Muratore, A. A. Voevodin, D. I. Koukis, D. B. Tanner, and D. J. Arenas, J. Appl. Phys. **114**, 083515 (2013).
[39]L. V. Rodríguez-de Marcos, J. I. Larruquert, J. A. Méndez, and J. A. Aznárez, Opt. Mater. Express **6**, 3622 (2016).
[40]D. E. Aspnes, S. M. Kelso, R. A. Logan, and R. Bhat, J. Appl. Phys. **60**, 754 (1986).
[41]A. V. Babichev, E. V. Nikitina, L. Ya. Karachinsky, I. I. Novikov, and A. Yu. Egorov, Proc. IEEE 266 (2024).
[42]A. A. Madigawa, M. S. Sultani Vala, and A. Demir, JPhys Photonics **7**, 045029 (2025).
[43]A. V. Babichev, Y. N. Kovach, S. A. Blokhin, L. Ya. Karachinsky, I. I. Novikov, A. Yu. Egorov, S.-C. Tian, and D. Bimberg, JPhys Photonics **7**, 032001 (2025).
[44]O. C. Karaman, G. N. Naidu, A. R. Bowman, E. N. Dayi, and G. Tagliabue, Laser Photonics Rev. **19**, e01014 (2025).
[45]T. Amotchkina, M. Trubetskov, D. Hahner, and V. Pervak, Appl. Opt. **59**, A40 (2019).
[46]L. V. Rodríguez-de Marcos, J. I. Larruquert, J. A. Méndez, and J. A. Aznárez, Opt. Mater. Express **7**, 989 (2017).
[47]L. V. Rodriguez de Marcos, M. Christophersen, M. F. Batkis, J. G. del Hoyo, J. Patel, E. J. Wollack, J. A. Christodoulides, J. I. Larruquert, and M. A. Quijada, Opt. Mater. Express **15**, 1750 (2025).
[48]M. Daimon, and A. Masumura, Appl. Opt. **41**, 5275 (2002).
[49]S. Spiga, W. Soenen, A. Andrejew, D. M. Schoke, X. Yin, J. Bauwelinck, G. Boehm, and M.-C. Amann, J. Lightwave Technol. **35**, 727 (2017).